\documentclass[preprint,12pt]{elsarticle}
\usepackage{graphicx,color,mathrsfs,amsmath,amssymb,amsthm,amsopn,bm}
\usepackage[nottoc]{tocbibind}      
\newlength{\lslashl}
\journal{Physics Letters B}                                                
\begin{document}

\begin{frontmatter}

\title{Photoproduction of hypernuclei within the quark-meson coupling model}
%\footnote{Authored by Jefferson
%Science Associates, LLC under U.S. DOE Contract No. DE-AC05-06OR23177.
%The U.S. Government retains a non-exclusive, paid-up, irrevocable, world-wide
%license to publish or reproduce this manuscript for U.S. Government purposes.
%}}

\author[JLab,Saha]{R. Shyam}
\author[EBAC]{K. Tsushima}
\author[JLab]{A. W. Thomas}

\address[JLab]{Theory Center, Thomas Jefferson National Accelerator Facility,
Newport News, VA 23606, USA}
\address[EBAC]{Excited Baryon Analysis Center and Theory Center,
Thomas Jefferson National Accelerator Facility, Newport News, VA 23606, USA}
\address[Saha]{Theory Division, Saha Institute of Nuclear Physics,  
Kolkata 700064, India }

\date{\today}

\begin{abstract}
We study the photoproduction of the $^{12}\!\!\!_\Lambda$B hypernucleus 
within a fully covariant effective Lagrangian based model, employing 
$\Lambda$ bound state spinors derived from the latest quark-meson coupling 
model. The kaon production vertex is described via creation, propagation and
decay of $N^*(1650)$, $N^*(1710)$, and $N^*(1720)$ intermediate baryonic 
resonant states in the initial collision of the photon with a target proton
in the incident channel. The parameters of the resonance vertices are fixed
by describing the total and differential cross section data on the elementary 
$\gamma p \to \Lambda K^+$ reaction in the energy regime relevant to the
hypernuclear production. It is found that the hypernuclear production 
cross sections calculated with the quark model based hyperon bound state
spinors differ significantly from those obtained with the phenomenological 
Dirac single particle wave functions.
\end{abstract}

\begin{keyword} 
Photoproduction of hypernuclei, covariant production model, quark-meson coupling
model hyperon spinors.
\PACS{21.80.+a, 13.60.-r, 13.75.Jz}
\end{keyword}

\end{frontmatter}

\newcommand{\vecr}{{\bf r}}
\newcommand{\veck}{{\bf k}}
\newcommand{\vecR}{{\bf R}}
\newcommand{\tauiso}{{\mbox{\boldmath $\tau$}}}
\newcommand{\gmu}{{\gamma_\mu}}
\newcommand{\gf}{{\gamma_5}}
\def\be{\begin{equation}}
\def\ee{\end{equation}}
\def\bg{\begin{eqnarray}}
\def\en{\end{eqnarray}}
\def\nn{\nonumber}
\def\ra{\rightarrow}
\def\la{\leftarrow}
%
%\begin{document}

Electromagnetic probes provide a very powerful tool for studying 
the $\Lambda$ hypernuclei. In contrast to the hadronic reactions 
[$(K^-,\pi^-)$ and $(\pi^+,K^+)$], a proton in the target nucleus 
is converted into a $\Lambda$ hyperon in both $(\gamma,K^+)$ and
$(e,e^\prime K^+)$ reactions, thus forming a neutron-rich hypernucleus.
This leads to the formation of mirror hypernuclear systems which can facilitate
the study of the charge symmetry breaking with strangeness degrees of freedom
(see, e.g.,~\cite{tam06,ban90,nem02,kei00}). Although in the electromagnetic 
reactions the momentum transfer to the nucleus is comparable to that of the 
$(\pi^+,K^+)$ reaction, they carry, in addition, significant spin-flip 
amplitudes due to the absorption of the photon spin and the forward angle 
domination of the cross sections. Furthermore, while the hadronic hypernuclear
production reactions are confined mostly to the nuclear surface because of
strong absorption of both $K^-$ and $\pi^\pm$, the electromagnetic reactions
occur deep inside the nucleus because of the weaker nuclear interactions of 
both photon and $K^+$. This makes them an ideal tool for studying deeply bound 
hypernuclear states.

Recently, Jefferson Laboratory (JLab) has started a systematic study of the
high-resolution hypernuclear production reactions on $p$-shell target nuclei
($^9$Be, $^{12}$C and $^{16}$O) using continuous electron beams 
\cite{miy03,yua06,iod07,cus08}. The quality and high resolution ($\sim$ 400 
keV) of the electron beam in these experiments make it possible to identify 
hyperon single particle states more clearly and to untangle the core 
excited states for the first time. While, the first measurements of  
hypernuclear production with real photons [($\gamma,K^+$) reaction] on a 
nuclear target ($^{12}$C) were reported long ago~\cite{yam95}, interest in 
this field has been revived with the possibility of performing more such 
measurements at accelerators MAMI-C in Mainz, and ELSA in BONN 
(see, e.g.~\cite{poc05}).

Several theoretical studies of photoproduction of hypernuclei have been 
reported~\cite{shi83,jos88,ben89,mot94,lee98,lee01}. They all use the 
framework of the impulse approximation, where the hypernuclear production 
amplitudes are calculated by determining expectation values of the operator
for the elementary $p (\gamma,K^+) \Lambda$ process. This operator is 
constructed either by using the Feynman diagrammatic approach including 
graphs corresponding to Born terms and resonance terms in $s$ and $u$ 
channels~\cite{shi83,jos88,lee01,ade85}, or phenomenologically by 
parameterizing the experimental cross sections for the elementary process
\cite{mot94,lee98}. Except for Ref.~\cite{ben89}, where Dirac spinors were 
used to describe the initial and final bound state wave functions, 
nonrelativistic models have been employed to obtain these wave functions 
in all of these investigations.  

On the other hand, in Ref.~\cite{shy08} a fully covariant model was employed 
to calculate the cross sections of the 
$^{16}O(\gamma,K^+){^{16}\!\!\!_{\Lambda}}N$ reaction. This model retains the 
full field theoretical structure of the interaction vertices and treats the 
baryons as Dirac particles (see also Ref.~\cite{shy04}). The initial
state interaction of the incoming photon with a bound proton leads to 
excitations of $N^*$(1650) [$\frac {1}{2}^-$], $N^*$(1710)[$\frac{1}{2}^+$],
and $N^*$(1720) [$\frac{3}{2}^+$] resonance intermediate states, which have 
been shown to make the predominant contributions to the $p(\gamma,K^+)\Lambda$
cross section~\cite{shk05}. In this model calculations are performed in 
momentum space throughout, hence it includes all the nonlocalities
in the production amplitudes that arise from the resonance propagators. 

However, the procedure of obtaining the bound state spinors in the 
previous application of this model brings in some uncertainty in the 
calculated hypernuclear production cross sections. In Ref.~\cite{shy08} 
the bound state spinors were computed in the coordinate space by   
solving the Dirac equation with scalar and vector fields having a 
Woods-Saxon radial form. With a set of radius and diffuseness parameters, 
the depths of these fields are searched so as to reproduce the binding 
energy (BE) of the given state. Because the experimental BEs of the 
hypernuclear states often involve ambiguities, the extracted potential 
depths also become ambiguous. Besides, the depths of the potential fields are 
dependent on the adopted radius and diffuseness parameters and there is no 
certain way of fixing them. Furthermore, both vector and scalar fields are 
assumed to have the same geometry.    

In this paper, we explore the feasibility of studying the photoproduction
of hypernuclei within the relativistic model of Ref.~\cite{shy08} but 
employing hyperon bound state spinors calculated within the quark-meson 
coupling (QMC) model. This provides an opportunity to investigate the role 
of the quark degrees of freedom in the hypernuclear production, which is 
a novel feature of this study. Since photoproduction of hypernuclei involves 
large momentum transfers~\cite{shy08e} to the target nucleus, it appears to 
be a good case for examining such short distance effects. 

In the QMC model~\cite{gui88,gui96,gui06,sai95}, quarks in the non-overlapping 
bags (modeled using MIT bag), interact self consistently with isoscalar-scalar
($\sigma$) and isoscalar-vector ($\omega$) mesons in the mean field 
approximation.  The explicit treatment of the nucleon internal structure 
represents an important departure from quantum hadro-dynamics (QHD) 
model~\cite{ser86}. The self-consistent response of the bound quarks to the 
mean $\sigma$ field leads to a new saturation mechanism for nuclear 
matter~\cite{gui88}. The QMC model has been used to study the properties of 
finite nuclei~\cite{sai96}, the binding of $\omega$, $\eta$, $\eta^\prime$ 
and $D$ nuclei \cite{tsu98a,tsu98b,tsu99,bas06} and also the effect of the 
medium on $K^\pm$ and $J/\Psi$ production~\cite{sai07}.

The most recent development of the quark-meson coupling model is the
inclusion of the self-consistent effect of the mean scalar field on the
familiar one-gluon exchange hyperfine interaction that in free space leads
to the $N-\Delta$ and $\Sigma-\Lambda$ mass splitting~\cite{rik07}. With
this~\cite{gui08} the QMC model has been able to explain the properties of 
$\Lambda$ hypernuclei for the $s$-states rather well, while the $p$- and
$d$-states tend to underbind.  It also leads to a very natural explanation 
of the small spin-orbit force in $\Lambda$-nucleus interaction. In this 
exploratory work, the bound $\Lambda$ spinors are generated from this version
of the QMC model and are used to calculate the cross sections of the 
$^{12}$C($\gamma,K^+)^{12}{\!\!\!_\Lambda}$B reaction. 
\begin{figure}[!t]
\begin{center}
\includegraphics[scale=0.45]{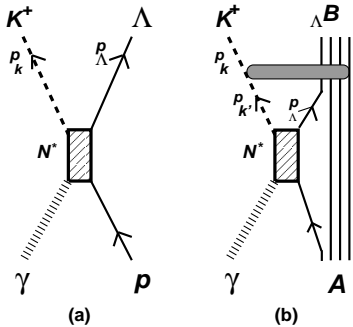}
\end{center}
\vspace{-0.5cm}
\caption{Representation of the type of Feynman diagrams included in our
calculations. The elliptic shaded area represents the optical model 
interactions in the outgoing channel.}
\label{fig_feyn}
\end{figure}

We fix the parameters of the resonance vertices by describing the data
on total and differential cross sections of the elementary $\gamma p \to 
K^+ \Lambda$ process in the relevant photon energy regime within a similar 
effective Lagrangian approach. This is in contrast to the calculations 
presented in Ref.~\cite{shy08} where they were taken from previous studies 
of photon and hadron induced associated $K^+\Lambda$ production reactions 
\cite{shk05,pen02,shy06}. Thus the resonance parameters used in the 
present study are better constrained.  

A preliminary experimental investigation of the $(\gamma, K^+)$ reaction on 
$^{12}$C was reported already in 1995~\cite{yam95}. Recently, the 
$^{12}{\!\!\!_\Lambda}$B  hypernucleus has been produced at JLab via the 
$(e,e^\prime K^+)$ reaction with a very high energy resolution
\cite{miy03,iod07}. In this experiment, apart from observing hypernuclear 
excitations where a proton is replaced (leaving $^{11}$B in the $3/2^-$ ground 
state) by a $\Lambda$ in $s$ and $p$ shells, one also sees identifiable 
strength in the region which corresponds to the excitation of the $^{11}$B 
core. This underlines the need of using a more microscopic hypernuclear 
structure model in describing the excitation of hypernuclear spectra in 
electromagnetic reactions. Our work is a first step in this direction where 
we examine the differences between the hypernuclear photoproduction cross 
sections obtained with a microscopic hypernuclear structure model and a 
phenomenological model. We restrict ourselves to photon energies below 1.5 
GeV as this is the relevant energy regime for the experiment performed 
already with real photons \cite{yam95}. Moreover, it has been shown in 
previous studies~\cite{ben89,lee98,shy08} that hypernuclear 
photoproduction cross sections on light targets peak around 0.95 - 1.0 GeV 
and drop off thereafter.    

As in the previous study~\cite{shy08}, we have used the graph of the type
shown in Fig.~1(b) to describe the hypernuclear production reaction
$A(\gamma,K^+){_{\Lambda}}B$. The elementary $\gamma p \to K^+ \Lambda$
process involved in this diagram is shown in Fig.~1(a). It is clear 
that our model has only $s$-channel resonance contributions. In principle, 
Born terms and resonance contributions in $u$- and $t$-channels should also
be included in the description of both the processes. These graphs constitute
the non-resonant background contributions. It should be noted that their 
magnitudes depend on the particular model used to calculate them and also 
on the parameters used within that model~\cite{mar06}. Except for photon 
energies close to threshold, these terms have been shown to make 
non-negligible contributions in the models of Refs.~\cite{shk05,mar06}. 
On the other hand, in Ref.~\cite{lee01} they have been found to be  
insignificant in both elementary as well as in-medium photon induced reactions
for beam energies below 1.5 GeV. We have ignored these diagrams in this 
exploratory work to keep our production model simple and similar to that of 
Ref.~\cite{shy08}. Furthermore, we reduce the computational complications 
further by using plane waves (PW) to describe the relative motion of the 
outgoing particle which is justified by the relatively weaker kaon-nucleus 
interaction in the final channel.
\begin{figure}[!t]
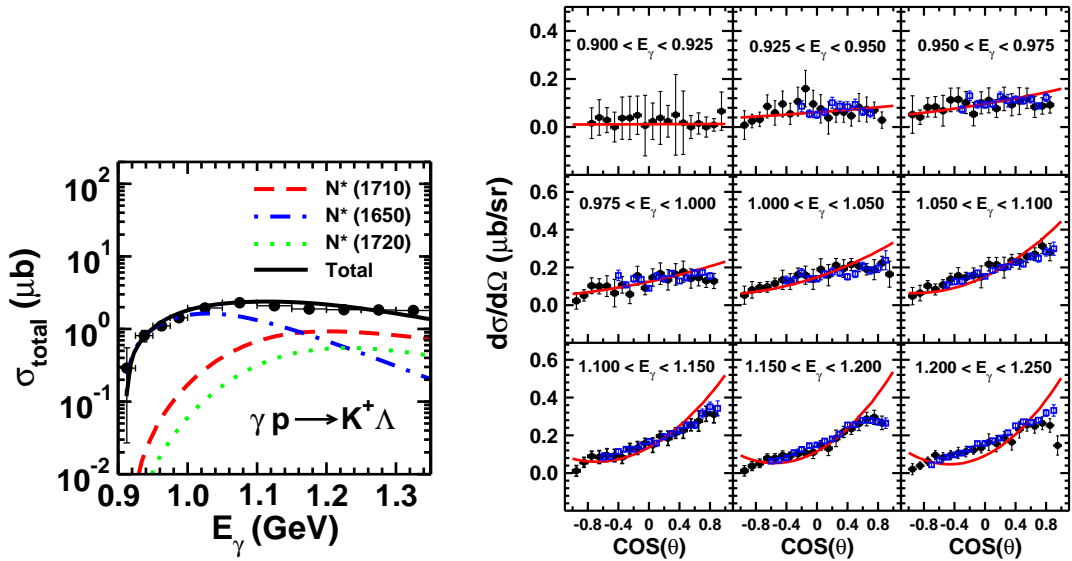

\begin{tabular}{cc}
\includegraphics[scale=0.55]{Fig2a.eps} & \hspace{0.10cm}
\includegraphics[scale=0.45]{Fig2b.eps} 
\end{tabular}
\caption{Calculated total cross sections (left panel) and differential 
cross section (right panel) for the $\gamma p \to \Lambda K^+$ reaction, 
obtained with the vertex constants of Table I as compared with the 
data from~\protect\cite{gla04} (solid circles) and~\protect\cite{bra06}
(open squares).} 
\label{fig_elem}
\end{figure}

All the ingredients (effective Lagrangians, resonance propagators etc.) 
required in calculations of the amplitudes associated with diagrams 1(a) 
and 1(b) are described in Refs.~\cite{shy08,shy04}. The coupling constants 
for the $N^*N\gamma$ ($g_{p\gamma}^{1,2}$) and $N^*K^+\Lambda$ ($g_{K\Lambda}$) 
vertices used in the present study (shown in table I) have been determined 
by comparing our calculations [graph 1(a)] with the total and differential 
cross section data for the elementary $\gamma p \to \Lambda K^+$ reaction 
in the relevant photon energy region. In Figs. 2(a) and 2(b) we have
shown the comparison of our results with the experimental data. While in 
Fig. 2(b) we have shown the data from both SAPHIR~\cite{gla04} and 
CLAS~\cite{bra06} collaborations, only SAPHIR data are shown in Fig.~2(a) 
as the CLAS total cross sections are somewhat uncertain due to absence of 
complete angular coverage. As was noted before~\cite{bra06}, in the energy 
range of our interest the CLAS and SAPHIR data agree with each other fairly 
well. Nevertheless, the CLAS data have much less statistical uncertainties. 
We see that calculated cross sections are in close agreement with the data 
in the considered photon energy regime. 

We further note that within our model $N^*(1650)$ resonance makes the dominant
contribution to the total cross section at lower photon energies, while 
$N^*(1710)$ is more important at higher energies. The contribution of 
$N^*(1720)$ is much weaker everywhere. We add however that this result is 
particular to our single channel model. In calculations where more resonances 
and the background terms are included, the pattern of relative resonance 
contributions could be different due to e.g. different kind of interference 
effects. Indeed the $N^*(1710)$ contribution has been found to be weak in 
several of the recent models~\cite{mar06,arn04,mar07}. In the unitary coupled 
channels calculations of Ref.~\cite{shk05} this resonance is suppressed 
because of the destructive interference with the background terms. However, 
since our purpose was to fix the parameters at the hyperon production 
vertices in the hypernuclear production reaction considered in Fig.~(2b), we 
consider the diagram 1(a) for the elementary reaction to be adequate. 
\begin{table}
\begin{center}
\caption{Resonances included in the calculations and their coupling
constants}
\begin{tabular}{cccccc}
\hline
Resonance & mass & width &$g_{p\gamma}^1$ &$g_{p\gamma}^2$ & $g_{K\Lambda}$\\
          & \footnotesize {GeV}&\footnotesize {GeV}& & &  \\
\hline
$N^*(1650)$ & 1.650 & 0.165 &-0.45 &      & 0.96 \\
$N^*(1710)$ & 1.710 & 0.180 & 0.25 &      &-6.10 \\
$N^*(1720)$ & 1.720 & 0.200 &-0.75 & 0.25 & 0.07 \\
\hline
\end{tabular}
\end{center}
\label{tab_respar}
\end{table}

The amplitudes for diagram 1(b) involve momentum space four component 
(spin space) Dirac spinors [$\psi$(p)] for bound nucleon and hyperon 
states~\cite{shy95} and the momentum space kaon-nucleus wave function 
[$\phi_{K}^{(-)*}(p_K^\prime, p_K)$] which can be calculated by using an 
appropriate $K^+$ - nucleus optical potential (see, {\it e.g.}, 
Ref.~\cite{tab77}). Momenta $p_K$ and $p_K^\prime$ are as defined  
in Fig.~1(b). In the PW approximation one writes 
$\Phi_{K}^{(-)*}(p_K^\prime, p_K)  =  \delta^4(p^\prime_K - p_K)$. 
 
The spinors, $\psi(p)$, are solutions of the Dirac equation in momentum 
space for a bound state problem in the presence of an external potential 
field~\cite{shy04,shy95}
\begin{eqnarray}
p\!\!\!/\psi(p) & = & m_N\psi(p) + F(p),
\end{eqnarray}
where
\begin{eqnarray}
F(p) & = & \delta(p_0 - E) \Biggl[\int d^3p^\prime V_s(-{\bf p}^\prime)
\psi({\bf p} + {\bf p}^\prime) \nonumber \\ & - & \gamma_0
 \int d^3p^\prime V_v^0(-{\bf p}^\prime) \psi({\bf p} + {\bf p}^\prime)
                   \Biggr] .
\end{eqnarray}
In Eq.~(2), the real scalar and timelike vector potentials $V_s$ and $V_v^0$
represent, respectively, the momentum space local Lorentz covariant 
interaction of single nucleon or $\Lambda$ with the remaining $(A-1)$ 
nucleons. We denote a four momentum by $p = (p_0,{\bf p}$). The magnitude 
of the three momentum ${\bf p}$ is represented by $k$, and its directions 
by ${\hat p}$. $p_0$ is the time like component of $p$. Spinors 
$\psi(p)$ and $F(p)$ are written as
\begin{eqnarray}
\psi(p) & = & \delta(p_0-E){{f(k) {\mathscr Y}_{\ell 1/2 j}^{m_j} (\hat p)
                   \choose {-ig(k)}{\mathscr Y}_{\ell^\prime 1/2 j}^{m_j}
                     (\hat p)}}, \nonumber \\
F(p) & = & \delta(p_0-E){{\zeta(k) {\mathscr Y}_{\ell 1/2 j}^{m_j} (\hat p)}
                \choose {-i\zeta^\prime(k){\mathscr Y}_{\ell^\prime 1/2 j}
                     ^{m_j} (\hat p)}},
\end{eqnarray}
where $f(k)$[$\zeta(k)$] is the radial part of the upper component
of the spinor $\psi(p)$[$F(p)$]. Similarly $g(k)$[$\zeta^\prime(k)$] are
the same of their lower component. $f(k)$ and $g(k)$ represent  
Fourier transforms of radial parts of the corresponding coordinate 
space spinors. $\zeta(k)$ are related to $f$, $g$ and the scalar and vector
potentials as shown in Ref.~\cite{shy04}. 
\begin{table}
\begin{center}
\caption{Depths of the Dirac vector ($V_v$) and scalar ($V_s$) fields for 
single particle $\Lambda$ and nucleon shells. In each case, both fields have 
the Woods-Saxon form with similar radius ($r$ = 0.983 fm) and diffuseness 
($a$ = 0.606 fm) parameters. Also shown are the experimental binding energies  
for each shell (numbers in the brackets are the BEs predicted by the QMC model).} 
\begin{tabular}{ccccc}
\hline 
 State & BE &$V_v$ &  $V_s$ \\
       &(\footnotesize{MeV})&(\footnotesize{MeV})& (\footnotesize{MeV}) \\
\hline
$^{12}\!\!\!_\Lambda$B$(1s_{1/2})$ & 11.37 (14.93) & 171.78 & -212.69 \\
$^{12}\!\!\!_\Lambda$B$(1p_{3/2})$ &  1.37 ( 3.62) & 204.16 & -252.28 \\
$^{12}\!\!\!_\Lambda$B$(1p_{1/2})$ &  1.03 ( 3.62) & 227.83 & -280.86 \\
$^{12}$C$(1p_{3/2})$               & 15.96         & 382.60 & -472.34 \\
\hline
\end{tabular}
\end{center}
\label{tab_bound}
\end{table}

In Table 2 we show the parameters associated with the scalar and vector 
fields of the phenomenological model for $\Lambda$ and nucleon bound states,
and the corresponding experimental BEs which are the averages of the values 
reported by several experimental studies~\cite{miy03,yua06,iod07,ahm03}.
In this table we also give the BEs of the $\Lambda$ bound states as predicted
by the QMC model.

To calculate the bound state spinors within the QMC model we have used its
latest version, where the calculations for $\Lambda$ and $\Xi$ hypernuclei 
are of comparable quality to earlier QMC results~\cite{tsu98b}. In addition, 
without requiring any  additional parameter it predicts no nuclear bound 
$\Sigma$ states~\cite{gui08}, which is in qualitative agreement with the 
experimental observations. This is facilitated by the extra repulsion 
associated with the increased one-gluon-exchange hyperfine in-medium 
interaction. We refer to Ref.~\cite{gui08} for more details of this new 
version of the QMC.

In order to calculate the properties of finite hypernuclei, we construct
a simple, relativistic shell model, with the nucleon core calculated in a 
combination of self-consistent scalar and vector mean fields. The 
Lagrangian density for a hypernuclear system in the QMC model is 
written as a sum of two terms, ${\mathscr L}^{HY}_{QMC}$ = ${\mathscr L}_{QMC}
+ {\mathscr L}^Y_{QMC}$, where 
\cite{tsu98a},  
%
%\begin{eqnarray}
%{\mathscr L}^{HY}_{QMC} &=& {\mathscr L}_{QMC} + {\mathscr L}^Y_{QMC},
%\nn \\
%
\begin{eqnarray}
&&\hspace{-1.4cm}{\mathscr L}_{QMC} =  \overline{\psi}_N(\vec{r})
[ i \gamma \cdot \partial - M_N(\sigma) - (\, g_\omega \omega(\vec{r}) \nn \\
&& + g_\rho \frac{\tau^N_3}{2}b(\vec{r})
+ \frac{e}{2} (1+\tau^N_3) A(\vec{r}) \,) \gamma_0 ] \psi_N(\vec{r}) \nn \\
%\quad \label{Lag1}\\ 
%
&& - \frac{1}{2}[ (\nabla \sigma(\vec{r}))^2 + m_{\sigma}^2
      \sigma(\vec{r})^2 ] \nn \\
&& + \frac{1}{2}[ (\nabla \omega(\vec{r}))^2 + m_{\omega}^2 \omega(\vec{r})^2 ]
\nn \\
&& + \frac{1}{2}[ (\nabla b(\vec{r}))^2 + m_{\rho}^2 b(\vec{r})^2 ] 
 +\frac{1}{2} (\nabla A(\vec{r}))^2, \label{Lag2} \\
&&\hspace{-1.4cm}{\mathscr L}^Y_{QMC} = \sum_{Y=\Lambda,\Sigma,\Xi} 
\overline{\psi}_Y(\vec{r}) [ i \gamma \cdot \partial - M_Y(\sigma)
- (\, g^Y_\omega \omega(\vec{r})\nn \\
&& + g^Y_\rho I^Y_3 b(\vec{r})
+ e Q_Y A(\vec{r}) \,) \gamma_0 ] \psi_Y(\vec{r}), \qquad \label{Lag3}
\end{eqnarray}
where $\psi_N(\vec{r})$ ($\psi_Y(\vec{r})$) and $b(\vec{r})$ are, respectively, 
the nucleon (hyperon) and the $\rho$ meson (the time component in the third 
direction of isospin) fields, while $m_\sigma$, $m_\omega$ and $m_{\rho}$ are
the masses of the $\sigma$, $\omega$ and $\rho$ mesons. $g_\omega$ and 
$g_{\rho}$ are the $\omega$-N and $\rho$-N coupling constants which are 
related to the corresponding (u,d)-quark-$\omega$, $g_\omega^q$, and $(u,d)$
quark-$\rho$, $g_\rho^q$, coupling constants as $g_\omega = 3 g_\omega^q$ and
$g_\rho = g_\rho^q$.  $I^Y_3$ and $Q_Y$ are the third component of the hyperon
isospin operator and its electric charge in units of the proton charge, $e$, 
respectively.

The following set of equations of motion are obtained for the hypernuclear 
system from the Lagrangian density Eqs.~(\ref{Lag2})-(\ref{Lag3}):
\begin{eqnarray}
&&\hspace{-1.4cm}[i\gamma \cdot \partial -M_N(\sigma)-
(\, g_\omega \omega(\vec{r}) + g_\rho \frac{\tau^N_3}{2} b(\vec{r}) \nn \\
&&\hspace{-1.4cm} + \frac{e}{2} (1+\tau^N_3) 
 A(\vec{r}) \,) \gamma_0 ] \psi_N(\vec{r}) =  0, \label{eqdiracn}\\
&&\hspace{-1.4cm}[i\gamma \cdot \partial - M_Y(\sigma)-
(\, g^Y_\omega \omega(\vec{r}) + g_\rho I^Y_3 b(\vec{r}) \nn \\
&&\hspace{-1.4cm} + e Q_Y A(\vec{r}) \,) \gamma_0 ] \psi_Y(\vec{r}) = 0, \label{eqdiracy}\\
%
%& &(-\nabla^2_r+m^2_\sigma)\sigma(\vec{r}) =
%- [\frac{\partial M_(\sigma)}{\partial \sigma}]\rho_s(\vec{r})
%- [\frac{\partial M_Y(\sigma)}{\partial \sigma}]\rho^Y_s(\vec{r}),
%\nn \\
&&\hspace{-1.4cm}(-\nabla^2_r+m^2_\sigma)\sigma(\vec{r})  = \nn \\
&&\hspace{-1.4cm} g_\sigma C_N(\sigma) \rho_s(\vec{r})
 + g^Y_\sigma C_Y(\sigma) \rho^Y_s(\vec{r}), \label{eqsigma}\\
&&\hspace{-1.4cm}(-\nabla^2_r+m^2_\omega) \omega(\vec{r})  =
g_\omega \rho_B(\vec{r}) + g^Y_\omega
\rho^Y_B(\vec{r}),\label{eqomega}\\
&&\hspace{-1.4cm}(-\nabla^2_r+m^2_\rho) b(\vec{r})  =
\frac{g_\rho}{2}\rho_3(\vec{r}) + g^Y_\rho I^Y_3 \rho^Y_B(\vec{r}),
 \label{eqrho}\\
&&\hspace{-1.4cm}(-\nabla^2_r) A(\vec{r})  =
e \rho_p(\vec{r})
+ e Q_Y \rho^Y_B(\vec{r}) ,\label{eqcoulomb}
\end{eqnarray}
where, $\rho_s(\vec{r})$ ($\rho^Y_s(\vec{r})$), $\rho_B(\vec{r})$
($\rho^Y_B(\vec{r})$), $\rho_3(\vec{r})$ and $\rho_p(\vec{r})$ are the 
scalar, baryon, third component of isovector, and proton densities at the 
position $\vec{r}$ in the hypernucleus~\cite{tsu98a}. On the right hand side
of Eq.~(\ref{eqsigma}), a new, and characteristic feature of QMC appears,  
arrising from the internal structure of nucleon and hyperon, namely,
$g_\sigma C_N(\sigma)= - [{\partial M_N(\sigma)}/{\partial \sigma}]$ 
and $g^Y_\sigma C_Y(\sigma)= - [{\partial M_Y(\sigma)}/{\partial \sigma}]$  
where $g_\sigma \equiv g_\sigma (\sigma=0)$ and $g^Y_\sigma \equiv 
g^Y_\sigma (\sigma=0)$. The scalar and vector fields as well as the spinors 
for hyperons and nucleons can be obtained by solving these coupled equations 
self-consistently.
\begin{figure}[!t]
\begin{center}
\includegraphics[scale=0.45]{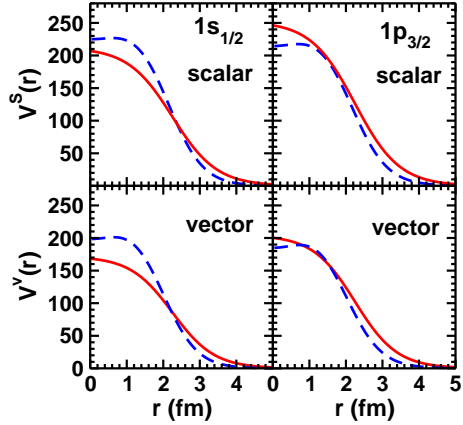}
\caption{ Scalar (upper panel) and vector (lower panel) 
potential fields for $1s_{1/2}$ and $1p_{3/2}$ $\Lambda$ states in 
$^{12}\!\!\!_\Lambda$B. The Dirac single particle and QMC model results are 
shown by solid and dashed lines, respectively.}  
\end{center}
\label{fig_pots}
\end{figure}

In Fig.~3, we compare the scalar and vector fields as calculated within the
QMC model with those of the phenomenological model for $1s_{1/2}$ and 
$1p_{3/2}$ $\Lambda$ states. It should be noted that in the QMC model the 
scalar and vector fields are generated by the couplings of the $\sigma$ and 
$\omega$ mesons to the quarks. Due to the different masses of these mesons and 
their couplings, especially the density dependence of the $\sigma N$ coupling 
strengths, the scalar and vector fields acquire different radial dependence. 
In contrast to this, the two fields have the same radial shapes in the 
phenomenological model. We further notice that for the $1s_{1/2}$ $\Lambda$ 
state the QMC scalar and vector fields are larger (smaller) in magnitude 
than those of the phenomenological model for $r < 2.5 ( > 2.5)$ fm. However, 
for the $1p_{3/2}$ state they are smaller than the phenomenological ones 
everywhere.
 
Fig.~4 shows the moduli of the upper and lower components of $1s_{1/2}$ and 
$1p_{3/2}$ $\Lambda$ hyperon spinors for the $^{12}\!\!\!_\Lambda$B in both 
coordinate space (upper panel) and momentum space (lower panel). We see that 
for the $1s_{1/2}$ $\Lambda$ bound state, the spinors of the QMC model differ
significantly from their phenomenological counterparts at both $r$ $< 2$ fm and
$r$ $> 4$ fm. For the $1p_{3/2}$ $\Lambda$ state while differences between 
them are quite big for $r$ $> 4$ fm, this is not as prominent at smaller radii.
On the other hand, in the momentum space the differences in the 
spinors of the two models are already quite large for $q > 1.0$ fm$^{-1}$ 
for the $1s_{1/2}$ state whereas for the $1p_{3/2}$ state the difference 
between the two becomes large for $q$  beyond 2  fm$^{-1}$. We also note   
that only for $q < 1.0$ fm$^{-1}$, is the magnitude of the lower
component ($|g(q)|$) substantially smaller than that of the upper component
($|f(q)|$). In the region of $q$ pertinent to the kaon production,
$|g(q)|$ may not be negligible. In fact, it has been shown
earlier~\cite{ben89} that the relativistic effects resulting from the small
component of Dirac bound states are large for the kaon photoproduction
reactions on nuclei.
\begin{figure}[!t]
\begin{center}
\includegraphics[scale=0.45]{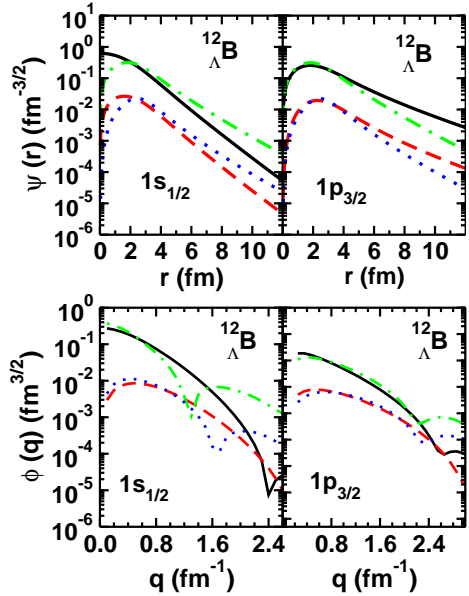}
\caption{ Moduli of upper($|f|$) and lower ($|g|$) components of the 
$1s_{1/2}$ and $1p_{3/2}$ $\Lambda$ orbits in $^{12}\!\!\!_\Lambda$B
hypernucleus in coordinate space (upper panel) as well as in  momentum space 
(lower panel). $|f|$ and $|g|$ of the phenomenological model are shown by 
solid and dashed lines, respectively while those of the QMC model by 
dashed-dotted and dotted lines, respectively.}
\end{center}
\label{fig_spinor}
\end{figure}

The threshold for the kaon photoproduction on $^{12}$C is about 695 MeV.
The momentum transfer involved in this reaction at 10$^\circ$  kaon
angle varies between approximately 2 fm$^{-1}$ to 1.4 fm$^{-1}$ in the
photon energy range of 0.7 GeV to 1.2 GeV~\cite{shy08e}. In Fig.~5, we 
compare the differential cross section obtained by using the
$\Lambda$ bound state spinors calculated within the QMC and the 
phenomenological models for the $^{12}$C$(\gamma,K^+)$$^{12}\!\!\!_\Lambda$B 
reaction. The hole state spinor was taken from the phenomenological model in
both cases. The cross sections are shown for photon energies in the
range of  0.7-1.2 GeV corresponding to the outgoing kaon 
angle of 10$^\circ$. The hypernuclear states populated are $1^-$, $2^-$, 
and $2^+$, $3^+$ corresponding to the  particle-hole configurations of  
$(1p_{3/2}^{-p},1s_{1/2}^\Lambda)$ and $(1p_{3/2}^{-p},1p_{3/2}^\Lambda)$, 
respectively. We see that in each case the QMC cross sections are smaller 
than those obtained with phenomenological hyperon spinors. For the 1$^-$ and 
2$^-$ states (involving $s$ state $\Lambda$ spinors), the QMC cross sections 
are lower because the corresponding momementum space spinors are smaller 
than their phenomenological model counterparts in the relevant momentum 
region. For the $2^+$ and $3^+$ states, additionally, the QMC potentials 
are also smaller than the phenomenological ones, which leads to lower 
QMC cross sections. In this figure we further note that the peaks of the 
QMC cross sections are somewhat shifted toward lower photon energies as 
compared to those of the phenomenological model. This can be understood 
from the fact that at lower photon energies the momentum transfer to the 
nucleus is relatively larger. In this region the QMC momentum space 
$\Lambda$ spinors are larger as compared to those of the phenomenological 
model
\begin{figure}[!t]
\begin{center}
\includegraphics[scale=0.45]{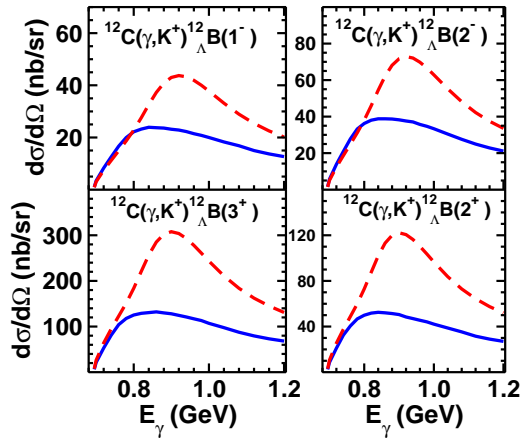}
\caption{ Differential cross sections (for the outgoing kaon 
angle of 10$^\circ$) for the $^{12}$C$(p,K^+)$$^{12}\!\!\!_\Lambda$B reaction 
leading to hypernuclear states as indicated. The solid and dashed lines show 
the results of QMC and phenomenological models, respectively.} 
\end{center}
\label{fig_cr}
\end{figure}

We further note that within each group the highest $J$ state is most 
strongly excited, which is in  line with the results presented in 
Refs.~\cite{ben89,shy08,ros88}. Furthermore, unnatural parity states within 
each group are preferentially excited by this reaction. The unnatural parity 
states are excited through the spin flip process. Thus this confirms 
that kaon photo- and also electro-production reactions on nuclei are ideal 
tools for investigating the structure of unnatural parity hypernuclear 
states. The addition of unnatural parity states to the spectrum of 
hypernuclei is expected to constrain the spin dependent part of the
effective $\Lambda-N$ interaction more tightly. 

In summary, we have studied the hypernuclear production by the $(\gamma,K^+)$ 
reaction on $^{12}$C within a covariant model, using hyperon
bound state spinors derived from the latest quark-meson coupling model. 
This is the first time that quark degrees of freedom has been explicitly 
invoked in the description of the hypernuclear production. In our model, 
in the initial collision of the photon with a target proton, $N^*(1710)$, 
$N^*(1650)$ and $N^*(1720)$ baryonic resonances are excited which subsequently
propagate and decay into a $\Lambda$ hyperon that gets captured in one of the
nuclear orbits, while the other decay product $K^+$ goes out. In contrast to 
the previous study within this model \cite{shy08}, we fix the coupling 
constants at both the electromagnetic and hadronic resonance vertices  
by describing both the total and the differential cross sections of the 
elementary $\gamma p \to \Lambda K^+$ reaction in the relevant region of 
photon energies. Thus the input parameters are better constrained in 
this study.

We have also performed calculations with bound $\Lambda$ spinors obtained
by solving the Dirac equation with vector and scalar potential fields having  
Woods-Saxon shapes. Their depths are fitted to the binding energies of the
respective states for a given set of geometry parameters which are taken to
be the same for the two fields. In contrast to this model, the QMC vector 
and scalar fields have different radial shapes. Furthermore, both shapes and
absolute magnitudes of the QMC fields are different from their Dirac 
counterparts. For the cases studied in this paper, the hypernuclear production
cross sections calculated with the QMC hyperon spinors and fields are not only
smaller in magnitude but also they peak at relatively lower photon energies as
compared to those obtained within the phenomenological model.
 
The distortion effects in the $K^+$ channel have not been included in
this study. However, as shown in Refs.~\cite{ben89,ros88}, these effects are 
weak for reactions on $p$-shell nuclei but they may be more significant for 
heavier systems. The cross sections as calculated in this paper may be 
uncertain to the extent of about 10$\%$ due to the non-inclusion of the 
nucleon intermediate states (Born terms).  

Our calculations further confirm that due to the selective excitation
of the high spin unnatural parity states, the $(\gamma,K^+)$ reaction on 
nuclei is an ideal tool for investigating the spin-flip transitions. 
Therefore, electromagnetic reactions provide a more complete knowledge of 
hypernuclear spectra and will impose more severe constraints on the 
poorly known spin dependent parts of the models of the $\Lambda-N$ 
interaction. Our model should be extended to electroproduction of 
hypernuclei (where the hadronic part remains the same as that discussed 
in this paper) so that the role of the quark degrees of freedom in the 
$\Lambda$ bound states can be checked against the data taken at JLab.

This work has been supported by the United States Department of Energy
contract no. DE-AC05-06OR23177 under which the Jefferson Science Associates
(JSA) operates the Thomas Jefferson National Accelerator Facility.

\end{document}